\documentclass[preprint,showpacs,preprintnumbers,amsmath,amssymb]{revtex4}

\usepackage{graphicx,color}

\begin{document}

\title{
Probing gravitational wave polarizations 
with Advanced LIGO, Advanced Virgo and KAGRA} 
\author{Yuki Hagihara}
\author{Naoya Era}
\author{Daisuke Iikawa}
\author{Hideki Asada} 
\affiliation{
Graduate School of Science and Technology, Hirosaki University,
Aomori 036-8561, Japan} 
\date{\today}

\begin{abstract} 
Assuming that, for a given source of gravitational waves (GWs), 
we know its sky position, as  
is the case of GW events with 
an electromagnetic counterpart such as GW170817, 
we discuss a null stream method to probe GW polarizations 
including spin-0 (scalar) GW modes and spin-1 (vector) modes, 
especially with an expected network of Advanced LIGO, 
Advanced Virgo and KAGRA. 
For two independent null streams for four  
non-co-aligned 
GW detectors, 
we study a location on the sky, 
exactly at which the spin-0 modes of GWs vanish 
in any null stream for the GW detector network, 
though the strain output at a detector may contain the spin-0 modes. 
Our numerical calculations show that there exist 
seventy 
sky positions 
that satisfy this condition of killing the spin-0 modes in the null streams. 
If a GW source with an electromagnetic counterpart is found 
in one of the 
seventy 
sky positions, 
the spin-1 modes will be testable separately from 
the spin-0 modes by the null stream method. 
In addition, we study a superposition of the two null streams 
to show that any one of the three modes 
(one combined spin-0 and two spin-1 modes) 
can be eliminated by suitably adjusting a weighted superposition of the null streams 
and thereby a set of the remaining polarization modes can be 
experimentally tested. 
\end{abstract}


\maketitle

\section{Introduction}
In Einstein's theory of general relativity (GR), gravitational waves (GWs) 
are spacetime perturbations that 
propagate at the speed of light 
and 
can be expressed in terms of the so-called ``tensor" modes 
which are spin-2 transverse and traceless (TT) 
\cite{Einstein1916, Einstein1918}. 
The propagation speed of GWs has been confirmed at the remarkable level 
$\sim O(10^{-15})$ 
by the celebrated GW170817 event with an electromagnetic counterpart 
that is the first observation of GWs from a NS-NS merger 
\cite{GW170817}. 
In the rest of this paper, we assume that the propagation speed of GWs 
is the same as the speed of light. 
In alternative theories to GR, specifically in  
general metric theories 
of gravity, 
GWs can contain extra degrees of freedom 
with spin 0 and spin 1, 
which are often called scalar and vector modes, respectively 
\cite{Eardley, PW}. 
Future detection of scalar or vector polarization would provide 
serious evidence against GR. 
Or future GW observations would put a stringent constraint on 
scalar and vector modes of GWs, which will lead to 
a new test of modified gravity theories, 
some of which might be excluded.  
Therefore, several attempts for 
polarization tests of GWs 
not only from bursts, compact binary coalescences, 
pulsars but also from stochastic sources have been discussed by 
many authors 
(e.g. \cite{Isi2015, Isi2017, Svidzinsky, pulsar, Nishizawa2009, Nishizawa2018}).

So far, aLIGO-Hanford (H), aLIGO-Livingston (L) and Virgo (V) 
have detected GW signals. 
However, three detectors are not enough 
for distinguishing polarizations of GWs. 
Very recently, Bruce Allen \cite{Allen} has examined whether  
the observation of the merging binary black hole event GW170814 
by the aLIGO-Virgo network 
\cite{GW170814} 
can be explained by 
a mimic model in which pure vector GWs mimic pure tensor one 
for the detector network of HLV. 
He shows that, if the vector polarizations in a hypothetical pure vector theory 
are allowed to rotate as the GWs propagate, then for a certain source location on the sky, 
the strain outputs of three GW detectors 
can reproduce the strain outputs that GR predicts. 
A kilometer-scale interferometer called KAGRA (K) 
is expected to join the GW detector network within a few years 
and to contribute as a fourth detector to break partly a degeneracy in polarization tests. 
See Reference \cite{LVK} for a comprehensive review on 
the expanding network of Advanced LIGO, Advanced Virgo and KAGRA 
(denoted as HLVK). 
Therefore, it is interesting to discuss how to probe 
GW polarizations by HLVK.

The main purpose of this paper is to study a possible method for probing 
GW polarizations including scalar and vector modes 
especially by taking account of the near-future network of HLVK, 
where we focus on GW events 
with an electromagnetic counterpart such as GW170817. 
For such a GW source, we know its location on the sky. 
In the rest of the paper, we use the information only on the direction 
of the GW source but not on the distance to the GW source. 
As a useful tool for our study, we consider null streams. 
A general method used in data analysis of GWs 
to separate signals from noise is that of null streams. 
The null stream approach was first introduced by G\"ursel and Tinto \cite{GT} 
and was extended later by Wen and Schutz \cite{WS} 
and Chatterji et al. \cite{null-papers}. 
The idea behind this method is that there exists 
a linear combination of the data from a network of detectors, 
which contains no tensor modes but only noise in GR cases. 
Chatziioannou, Yunes and Cornish have argued null streams 
for six (or more) GW detectors to probe GW polarizations 
\cite{CYC}. 

This paper is organized as follows. 
In Section II, we describe null streams 
for four 
non-co-aligned 
detectors and it is shown, 
by explicit calculations, that there are two independent 
null streams.  
By using two independent null streams, in Section III, 
we study a certain source location on the sky, 
exactly at which the null streams contain no spin-0 GW modes. 
We examine also superpositions of null streams for probing GW polarizations. 
Section IV is devoted to the Conclusion. 
Throughout this paper, 
Latin indices $a, b, \cdots$ run from 1 to 4 
corresponding to four detectors.

\section{Null streams for four non-co-aligned detectors}
Let us assume that there exist four detectors with uncorrelated noise 
and that, for a given source,  we know its sky position, 
as is 
the case of GW events with an electromagnetic counterpart such as GW170817. 
Then, one knows exactly how to shift the arrival time of the GW 
from detector to detector. 

For a detector labeled by $a$ ($a=1,2,3$ and $4$), 
the noise-weighted signal from a GW source 
at location denoted as $(\theta, \phi)$ on the sky is
\begin{align}
S_a =& 
F_a^{+} h^{+} + F_a^{\times} h^{\times} 
\nonumber\\
&+ F_a^{S} h^S + F_a^{L} h^L 
\nonumber\\
&+ F_a^{V} h^V + F_a^{W} h^W + n_a ,
\label{S0}
\end{align}
where 
$n_a$ denotes noise,  
$h^{+}$ and $h^{\times}$ denote the spin-2 modes called 
the plus and cross mode, respectively, 
$h^S$ and $h^L$ denote the  
spin-0 
modes called 
the breathing and longitudinal mode, respectively, 
and 
$h^V$ and $h^W$ denote the spin-1 modes 
often called the vector-$x$ and vector-$y$ mode, respectively 
\cite{footnote0} 
and 
$F_a^{+}$, $F_a^{\times}$, $F_a^{S}$, $F_a^{L}$, 
$F_a^{V}$ and $F_a^{W}$ 
are the antenna patterns for polarizations of GWs 
\cite{PW,Nishizawa2009,ST}. 
The antenna patterns are functions of 
a GW source location  
$\theta$ and $\phi$ \cite{footnote1}. 
In our numerical calculations, 
$\theta$ and $\phi$ denote the 
latitude and longitude, respectively. 

By noting $F_a^{S} = - F_a^{L}$ 
\cite{footnote2}
that was shown by 
Nishizawa et al. in \cite{LVK}, 
Eq. (\ref{S0}) can be simplified as 
\begin{align}
S_a =& 
F_a^+ h^{+} + F_a^{\times} h^{\times} 
\nonumber\\
&+ F_a^S (h^S - h^L) 
\nonumber\\
&+ F_a^V h^V + F_a^W h^W + n_a . 
\label{S}
\end{align}
Note that the effects of $h^S$ on the detector 
are exactly the same with the opposite sign as those of $h^L$. 
Hence, only the difference as $h^S - h^L$ can be tested 
as one combined spin-0 mode.

In this and next paragraphs of the present paper, 
we follow G\"ursel and Tinto \cite{GT} 
to consider only the purely tensorial modes $h^{\times}$ and $h^{+}$. 
In addition, let us imagine, for its simplicity, an ideal case that 
noise is negligible in Eq. (\ref{S}). 
By eliminating the two TT modes in signals at three detectors in the ideal case, 
we obtain a null stream \cite{GT} as, for $a=1, 2$ and $3$ for instance, 
\begin{align}
\delta_{23}S_1+\delta_{31}S_2+\delta_{12}S_3 = 0 ,  
\label{Null-123}
\end{align}
where we define 
\begin{align}
\delta_{ab} \equiv F_a^+ F_b^{\times} - F_b^+ F_a^{\times} . 
\end{align}

\begin{figure}
\includegraphics[width=16cm]{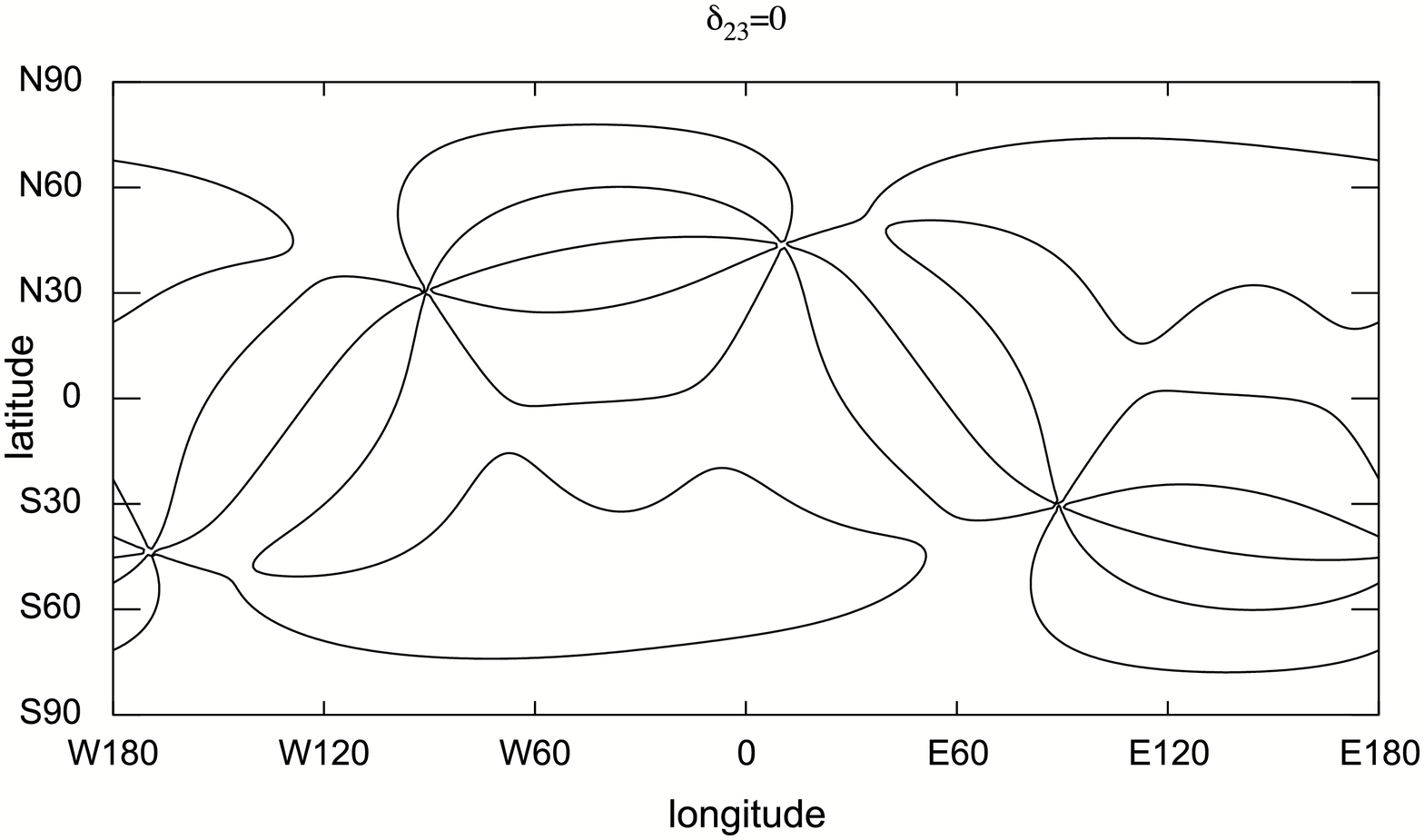}
\caption{
Curves for $\delta_{23} = 0$ in the sky, 
where L=2 and V=3 are assumed. 
}
\label{figure-delta23}
\end{figure}

Similarly, the other null streams are obtained as 
\begin{align}
&\delta_{34}S_2+\delta_{42}S_3+\delta_{23}S_4 = 0 , 
\label{Null-234}
\\
&\delta_{41}S_3+\delta_{13}S_4+\delta_{34}S_1 = 0 , 
\label{Null-341}
\\
&\delta_{12}S_4+\delta_{24}S_1+\delta_{41}S_2 = 0 . 
\label{Null-412}
\end{align}
By explicit calculations, we shall show that 
any two out of the null streams Eqs. (\ref{Null-123}) and (\ref{Null-234})-(\ref{Null-412}) 
can construct the remaining two null streams. 
Note that Reference \cite{null-papers} suggests that 
two null streams are likely to be enough for four 
non-co-aligned 
detectors. 
Without loss of generality, 
we take Eqs. (\ref{Null-123}) and (\ref{Null-234}) 
as two null streams and 
multiply them 
by $\delta_{34}$ and $\delta_{31}$, respectively. 
The difference between them is rearranged as 
\begin{align}
\delta_{23} (\delta_{41}S_3+\delta_{13}S_4+\delta_{34}S_1) = 0 , 
\label{proof1}
\end{align}
where we use an identity as 
\begin{align}
\delta_{12}\delta_{34} - \delta_{42}\delta_{31} = \delta_{23} \delta_{41} . 
\label{identity}
\end{align}
We thus obtain Eq. (\ref{Null-341}). 
Note that $\delta_{23}$ is nonvanishing almost everywhere 
in the sky for the 
non-co-aligned 
$a=2$ and $3$ detectors. 
See Figure \ref{figure-delta23} 
for the zero points of $\delta_{23}$ in the sky, 
where we assume a case of L=2 and V=3. 
In our numerical calculations, the reference for polarization angles 
is chosen as aLIGO-Livingston (L=2) and 
we assume that the source is located sufficiently far from the detectors, 
so that the plane wave approximation of GWs can be used. 
In the same way, we can obtain Eq. (\ref{Null-412}) from 
Eqs. (\ref{Null-123}) and (\ref{Null-234}) almost everywhere in the sky. 

Actually, a real GW detector has noise. 
Eqs. (\ref{Null-123}), (\ref{Null-234}), (\ref{Null-341}) 
and (\ref{Null-412}) are thus modified with noise as 
\begin{align}
\delta_{23}S_1+\delta_{31}S_2+\delta_{12}S_3 
&= 
\delta_{23}n_1+\delta_{31}n_2+\delta_{12}n_3 , 
\label{Null-123-n}
\\
\delta_{34}S_2+\delta_{42}S_3+\delta_{23}S_4 
&= 
\delta_{34}n_2+\delta_{42}n_3+\delta_{23}n_4 , 
\label{Null-234-n}
\\
\delta_{41}S_3+\delta_{13}S_4+\delta_{34}S_1 
&= 
\delta_{41}n_3+\delta_{13}n_4+\delta_{34}n_1 , 
\label{Null-341-n}
\\
\delta_{12}S_4+\delta_{24}S_1+\delta_{41}S_2 
&= 
\delta_{12}n_4+\delta_{24}n_1+\delta_{41}n_2  . 
\label{Null-412-n}
\end{align}
By using identities such as Eq. (\ref{identity}) again, it is clear 
that any pair among Eqs. (\ref{Null-123-n})-(\ref{Null-412-n}) 
can derive the remaining two equations.

\begin{figure}
\includegraphics[width=9.5cm]{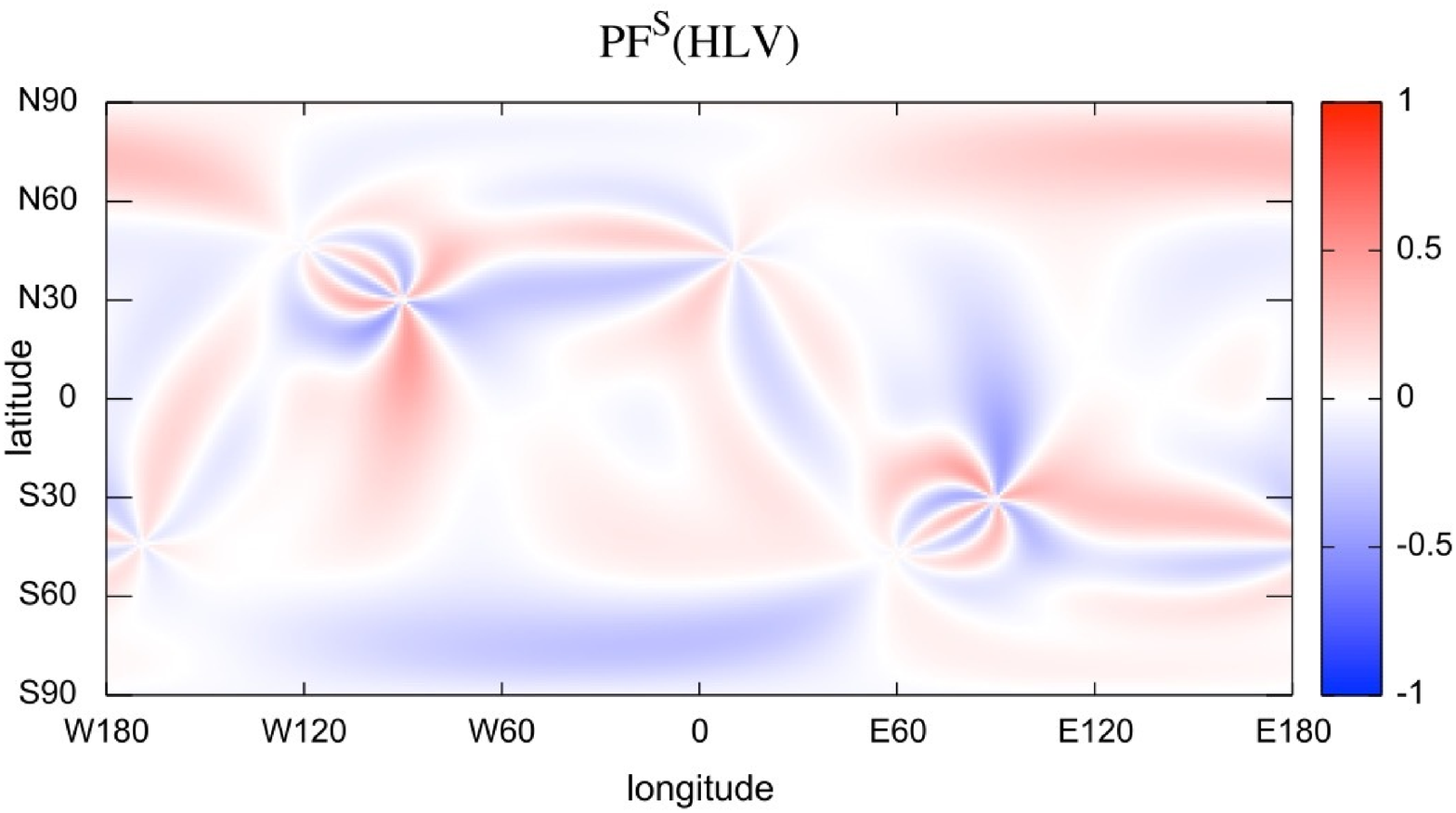}
\includegraphics[width=9.5cm]{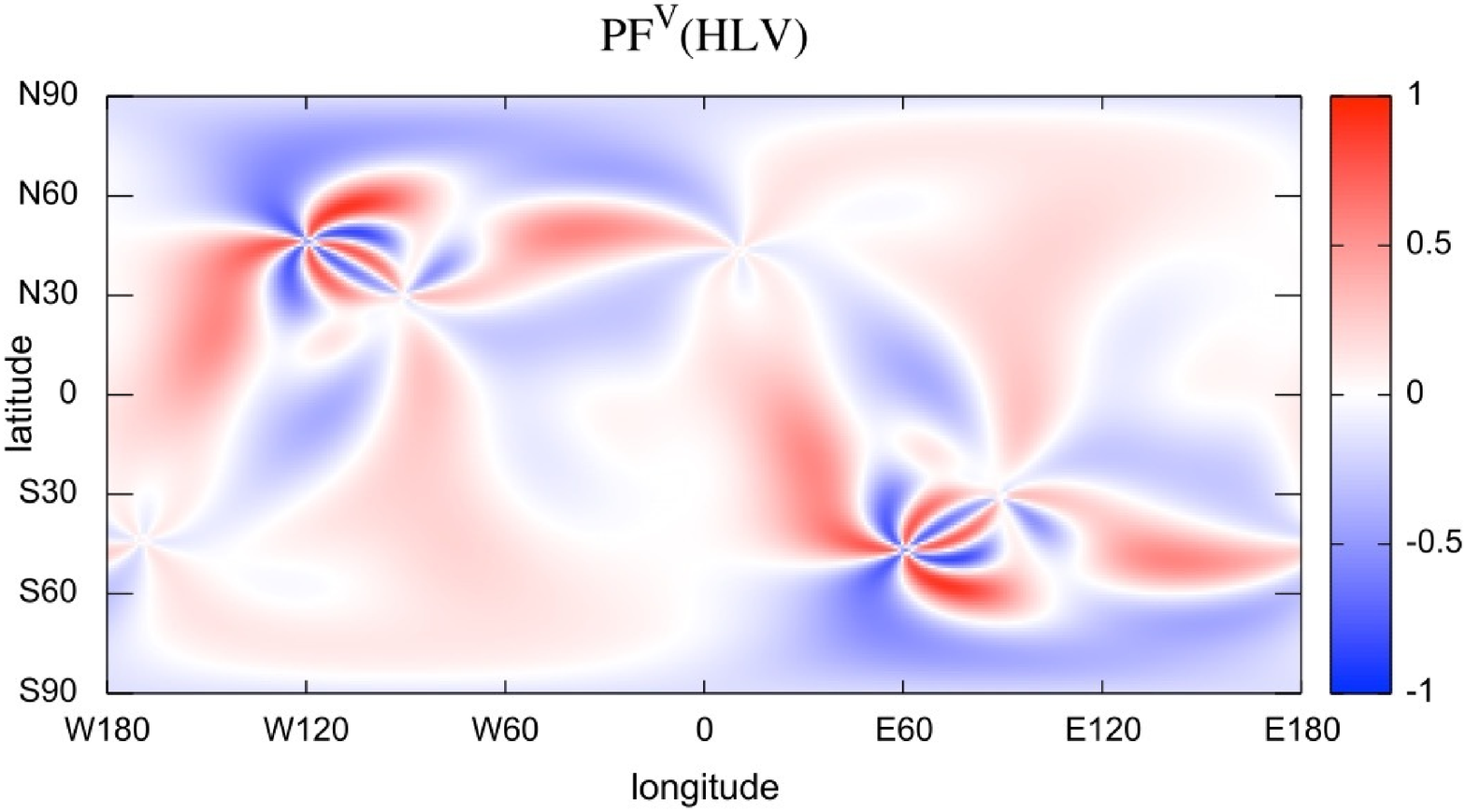}
\includegraphics[width=9.5cm]{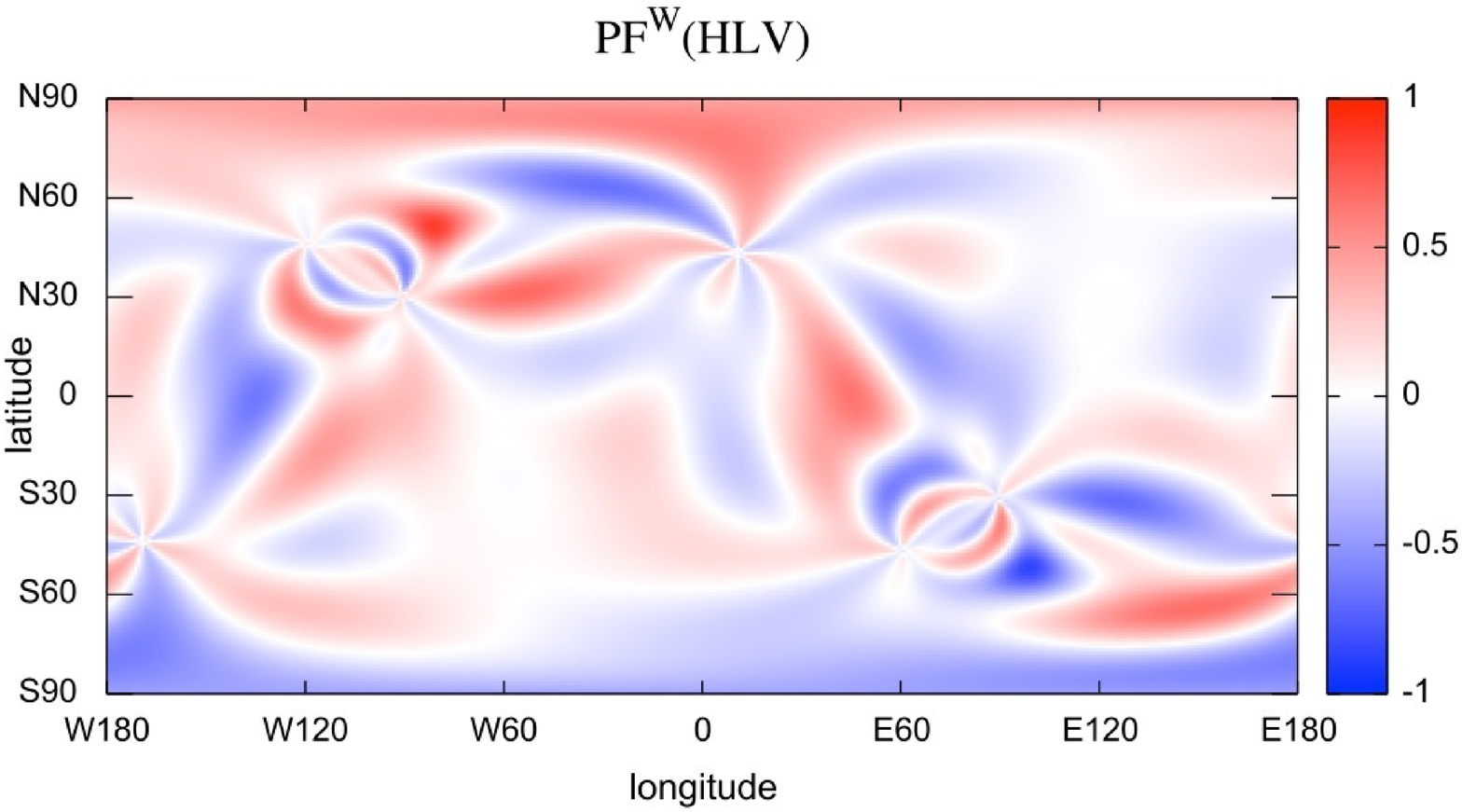}
\caption{
Contour map of the coefficients in the null stream 
as Eq. (\ref{PS}), 
where $\theta$ and $\phi$ denote 
the latitude and longitude, respectively. 
The detectors are labeled as H=1, L=2, V=3 and K=4. 
In this figure, white (in color) means zero, 
red (in color) is positive and 
blue (in color) denotes negative. 
From top to bottom: 
$P_a
F_a^S$, $P_b
F_b^V$ and $P_c
F_c^W$. 
}
\label{figure-contour1}
\end{figure}

\begin{figure}
\includegraphics[width=9.5cm]{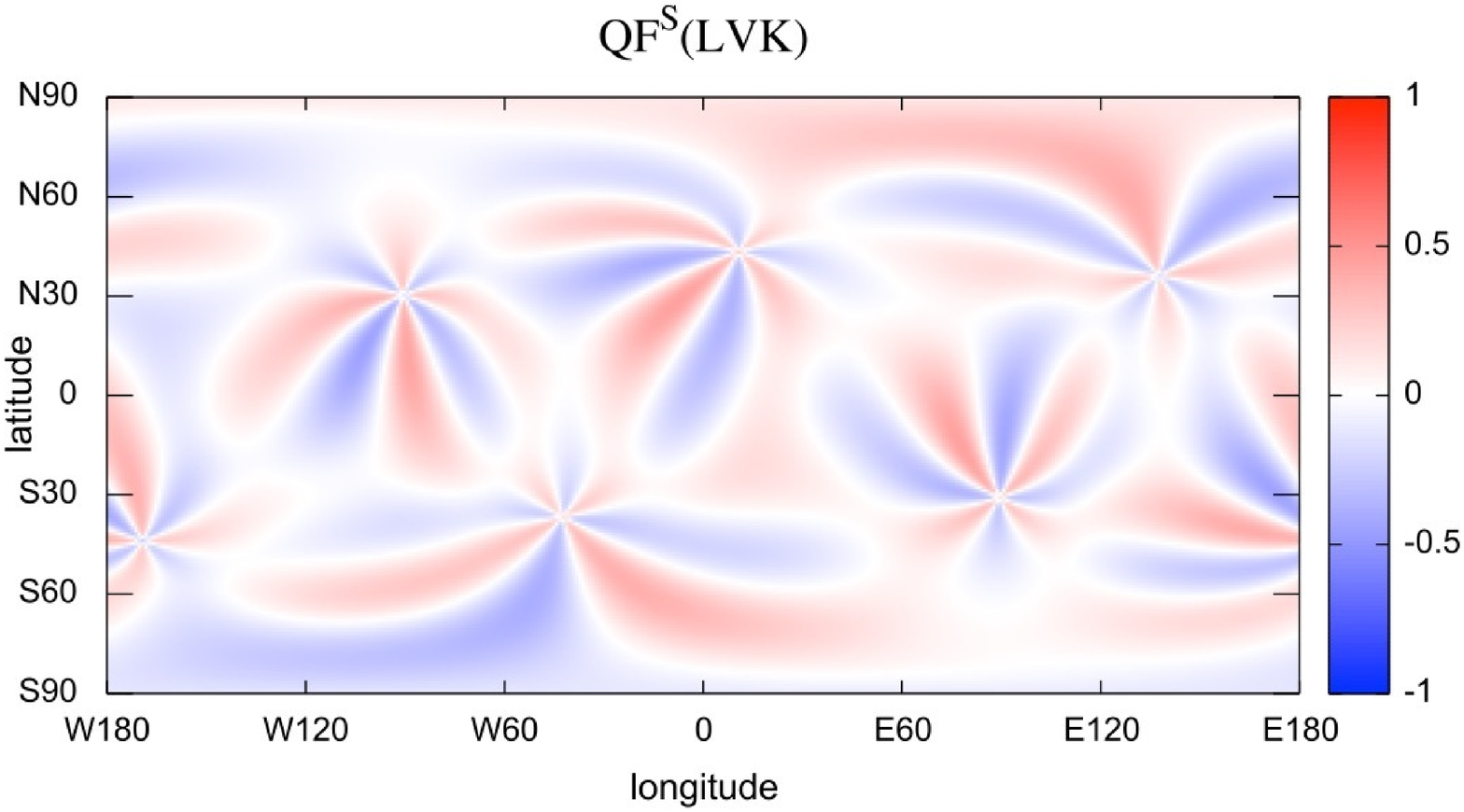}
\includegraphics[width=9.5cm]{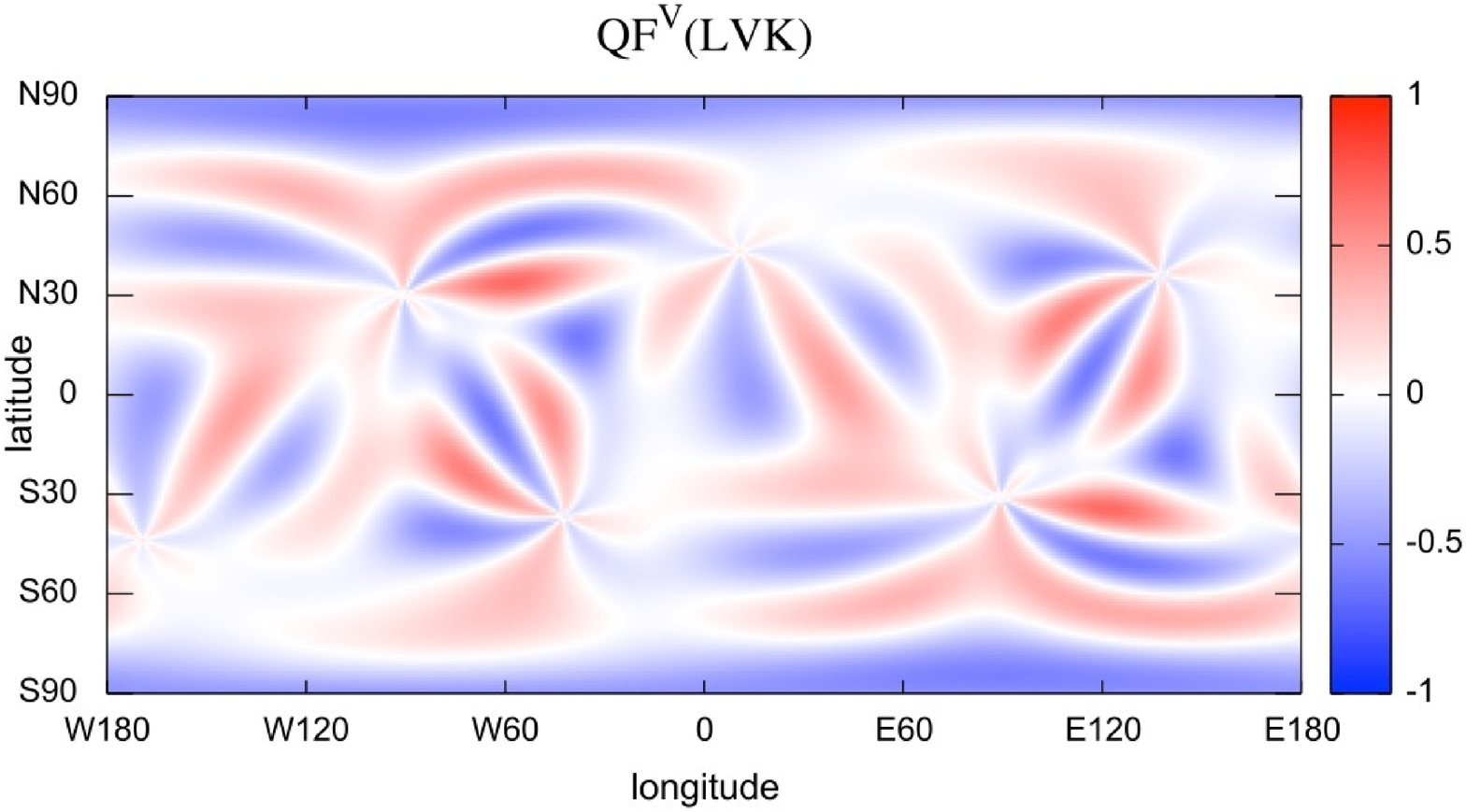}
\includegraphics[width=9.5cm]{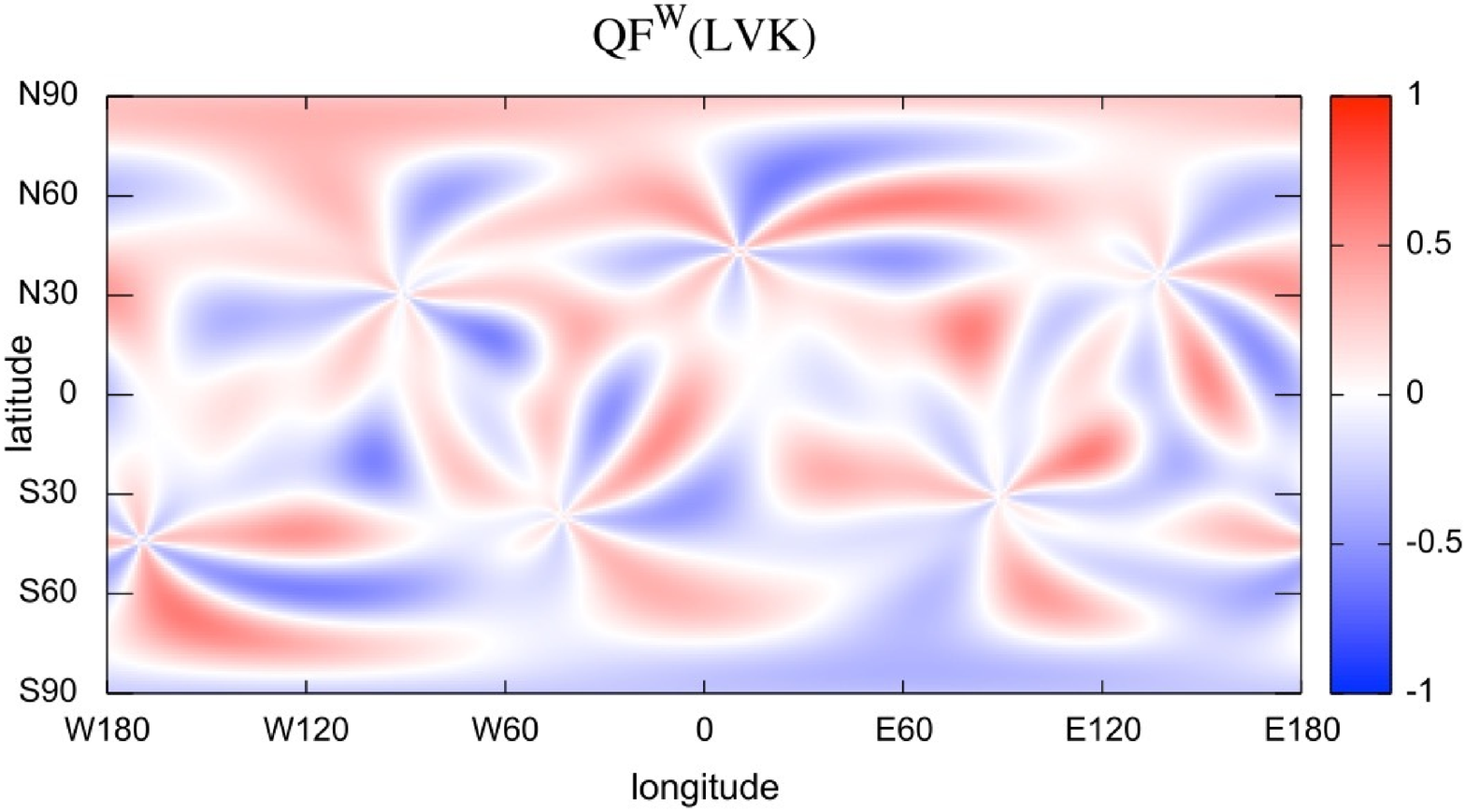}
\caption{
Contour map of the coefficients in the null stream 
as Eq. (\ref{QS}). 
This corresponds to Figure \ref{figure-contour1}. 
From top to bottom: 
$Q_a
F_a^S$, $Q_b
F_b^V$ and $Q_c
F_c^W$. 
}
\label{figure-contour2}
\end{figure}

Next, we incorporate scalar and vector polarization modes. 
Let us denote two null streams including spin-0 and spin-1 polarizations as 
\begin{align}
P_a S_a &= (P_bF_b^S)(h^S - h^L) + (P_cF_c^V)h_V + (P_dF_d^W)h_W 
+ P_e n_e ,
\label{PS}
\\
Q_f S_f &= (Q_gF_g^S)(h^S - h^L) + (Q_hF_h^V)h_V + (Q_iF_i^W)h_W 
+ Q_j n_j , 
\label{QS}
\end{align}
where we use Eq. (\ref{S}) and 
the summation is taken over $a, \cdots, j=1, 2, 3$ and $4$. 
Note that the tensor null stream is built in and hence $h^{+}$ 
and $h^{\times}$ do not appear in the above equations. 
Without loss of generality, we can choose 
$P_a$ and $Q_a$ as 
$(P_a) = (\delta_{23}, \delta_{31}, \delta_{12}, 0)$ and 
$(Q_a) = (0, \delta_{34}, \delta_{42}, \delta_{23})$ 
for its simplicity, 
which are corresponding to Eqs. (\ref{Null-123-n}) and (\ref{Null-234-n}) 
in the previous paragraph. 
In the next section, we shall examine Eqs. (\ref{PS}) and (\ref{QS}) 
in more detail. 
In numerical calculations for the HLVK network, 
we choose H=1, L=2, V=3 and K=4 
for $a=1, 2, 3$ and $4$ 
for its simplicity. 
See Figures \ref{figure-contour1} and \ref{figure-contour2} 
for the network of HLVK. 

\section{Spin-0 silent location on the sky and canceling method}
\subsection{Source locations on the sky at which null streams contain no spin-0 modes} 
Let us examine spin-0 modes in the null streams 
by Eqs. (\ref{PS}) and (\ref{QS}), in which 
the coefficients in front of $h^S - h^L$ are $P_a
F_a^S$ and $Q_a
F_a^S$. 
Note that the tensor null stream is built in and hence $h^{+}$ 
and $h^{\times}$ do not appear in the equations. 
If $P_a
F_a^S$ and $Q_a
F_a^S$ vanish simultaneously, 
spin-0 GW components do not contribute to the null streams, 
even if these waves had a nonnegligible amplitude. 
For a source location satisfying simultaneously $P_a
F_a^S = 0$ and $Q_a
F_a^S = 0$, therefore, 
the null streams can be used for testing only the vector modes 
$h^V$ and $h^W$. 

\begin{figure}
\includegraphics[width=16cm]{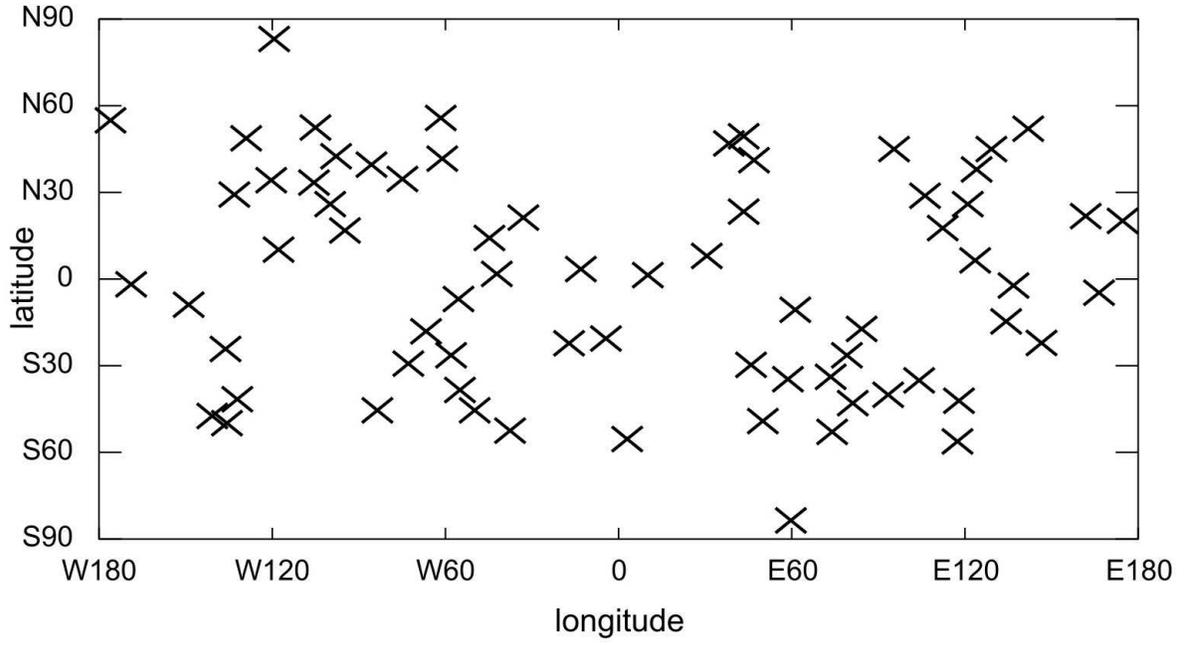}
\caption{
The 
seventy 
sky positions that satisfy simultaneously 
$P_a
F_a^S = 0$ and $Q_a
F_a^S = 0$, 
where we assume H=1, L=2, V=3 and K=4. 
}
\label{figure-zeros}
\end{figure}

As Section II suggests, 
$h^S-h^L$ for this very particular case will disappear 
in any null stream that is made of 
Eqs. (\ref{Null-123}), (\ref{Null-234}), (\ref{Null-341}) and (\ref{Null-412}), 
though the strain output at a detector may contain spin-0 modes. 
This can be explicitly shown as follows. 
Without loss of generality, we can consider Eq. (\ref{Null-341-n}) 
as another null stream, for which we define 
$(R_a) = (\delta_{34}, 0, \delta_{41}, \delta_{13})$. 
The coefficient of $h^S - h^L$ in this null stream is 
written as $R_a
F_a^S$. 
By using the same method that we use 
for obtaining Eq. (\ref{proof1}),  
one can find an identity as  
\begin{align}
\delta_{34}(P_aF_a^S) - \delta_{31}(Q_bF_b^S) 
= \delta_{23} (R_cF_c^S) , 
\end{align}
where we use Eq. (\ref{identity}). 
Here, $\delta_{23}$ is nonvanishing almost everywhere for 
non-co-aligned 
detectors. 
Therefore, if $P_a
F_a^S=0$ and $Q_b
F_b^S=0$ are satisfied simultaneously, 
$R_c
F_c^S$ vanishes. 
See Figure \ref{figure-zeros} for numerical plots 
of the sky locations that satisfy both $P_a
F_a^S=0$ and $Q_b
F_b^S=0$ 
for the HLVK network. 
There are 
seventy 
sky positions that kill the spin-0 modes 
in the null streams for HLVK. 

Why is the number of the specific sky positions seventy 
$\sim O(10^2)$ but not $O(1)$ nor $O(10^4)$? 
Here, let us discuss it from a mathematical viewpoint. 
Both $P_a F_a^S$ and $Q_a F_a^S$ are expressed in terms of 
trigonometric functions of $\theta$ and $\phi$ 
of degree twelve. 
Roughly speaking, they can be approximated by polynomials 
with $O(10)$ degrees. 
B\'ezout's theorem in algebraic geometry 
\cite{Bezout} 
states that two algebraic curves of degrees $m$ and $n$ on a plane 
intersect in $m \times n$ points if each point is counted 
with its intersection multiplicity. 
In the current paper, we do not count the intersection multiplicity. 
In our case, therefore, B\'ezout's theorem gives the upper bound on 
the number of the intersection points. 
The number of the specific sky positions is thus 
below $O(10) \times O(10) = O(10^2)$. 
Therefore, seventy points seem to be reasonable.

If we are extremely lucky to observe such a GW event 
with an electromagnetic counterpart 
at the location at which spin-0 modes fade out from the null streams, 
Eqs. (\ref{PS}) and (\ref{QS}) will enable to constrain 
(or perhaps detect) $h^V$ and $h^W$, separately. 
They are solved for $h^V$ and $h^W$ as 
\begin{align}
\left(
\begin{array}{c}
h^V \\
h^W 
\end{array}
\right)
&= 
\left(
\begin{array}{cc}
P_a 
F_a^V & P_b 
F_b^W\\
Q_c 
F_c^V & Q_d 
F_d^W
\end{array}
\right)^{-1} 
\left(
\begin{array}{c}
P_e (S_e - n_e) \\
Q_f (S_f - n_f) 
\end{array}
\right) , 
\label{vector}
\end{align}
where $a, \cdots, f$ run from $1$ to $4$. 
By observations using the four detectors, 
Eq. (\ref{vector}) can determine, in principle, 
$h^V$ and $h^W$ separately. 
In practice, we have noise in the measurements, 
so that we would be able to put a constraint on 
$h^V$ and $h^W$ within the observational error bars.

\subsection{How to adjust a weighted superposition of two null streams}
We consider a superposition of the two null streams by Eqs. (\ref{PS}) and (\ref{QS}) as 
\begin{align}
A P_aS_a + B Q_b S_b 
=& 
\left(A P_c
F_c^S + B Q_d
F_d^S\right) 
(h^S-h^L) 
\nonumber\\
&
+ \left(A P_e
F_e^V + B Q_f
F_f^V\right) 
h^V 
\nonumber\\
&
+ \left(A P_g
F_g^W + B Q_h
F_h^W\right) 
h^W 
\nonumber\\
&
+ \left(A P_i + B Q_i\right) n_i , 
\label{superposition}
\end{align}
where $A$ and $B$ are real constants. 
Note that the tensor null stream is built in and hence $h^{+}$ 
and $h^{\times}$ do not appear in the above equations. 
For the later convenience, 
let us denote a ratio between $A$ and $B$ as 
\begin{align}
r \equiv \frac{B}{A} . 
\label{ratio}
\end{align}

First, we consider a case that 
$A P_c
F_c^S + B Q_d
F_d^S$ in Eq. (\ref{superposition}) vanishes, 
though the strain at a detector may contain spin-0 modes. 
This case is achieved by choosing a ratio between $A$ and $B$ as 
\begin{align}
r^{S-L} 
\equiv
- \frac{P_a
F_a^S}{Q_b
F_b^S} ,
\label{ratio-S}
\end{align}
where we assume $Q_b
F_b^S \neq 0$. 
If $Q_b
F_b^S = 0$, Eq. (\ref{QS}) has no spin-0 parts, so that 
Eq. (\ref{QS}) can be used to probe the spin-1 modes as it is. 
For example, we consider two simple cases 
as GWs from the direction of the north pole (N) 
and GWs from the direction of the equator and prime meridian (EPM). 
For these cases, the ratio for the HLVK network is 
$r^{S-L}_N = -0.33669$ 
and 
$r^{S-L}_{EPM} = 0.31524$, 
respectively.

Next, we examine a case that the coefficient in front of $h^V$ 
in Eq. (\ref{superposition}) vanishes. 
This case can be achieved by choosing the ratio $r$ as 
\begin{align}
r^V 
\equiv 
- \frac{P_a
F_a^V}{Q_b
F_b^V} ,
\label{ratio-V}
\end{align}
where we assume $Q_b
F_b^V \neq 0$. 
If $Q_b
F_b^V$ vanishes, Eq. (\ref{QS}) does not contain $h^V$, 
such that 
Eq. (\ref{QS}) can immediately constrain some combination of 
$h^S-h^L$ and $h^W$. 
The ratio for HLVK is 
$r^{V}_N = -0.31112$ 
for GWs from the direction of the north pole 
and  
$r^{V}_{EPM} = 0.49763$
for GWs from the EPM direction.

Thirdly, we study a case that the coefficient of $h^W$ vanishes 
in Eq. (\ref{superposition}). 
This case can be achieved by choosing the ratio as 
\begin{align}
r^W 
\equiv 
- \frac{P_a
F_a^W}{Q_b
F_b^W} ,
\label{ratio-W}
\end{align}
where we assume $Q_b
F_b^W \neq 0$. 
If $Q_b 
F_b^W=0$, $h^W$ disappears in Eq. (\ref{QS}), 
so that 
Eq. (\ref{QS}) can immediately constrain some combination of 
$h^S-h^L$ and $h^W$. 
For HLVK, the ratio is  
$r^{W}_N = -1.61958$  
and 
$r^{W}_{EPM} = 0.54243$ 
for GWs from the N direction and from the EPM direction, respectively.

\section{Conclusion} 
In expectation of the near-future network of Advanced LIGO, 
Advanced Virgo and KAGRA, 
we discussed a null stream method to probe GW polarizations 
including spin-0 (scalar) GW modes and spin-1 (vector) modes, 
where we assumed that, for a given source of GWs, 
we know its sky position, as 
is the 
case of GW events with 
an electromagnetic counterpart such as GW170817. 
We studied a location on the sky, 
exactly at which the spin-0 modes of GWs vanish 
in null streams for the GW detector network, 
though the strain output at a detector may contain the spin-0 modes. 
By numerical calculations, we showed that there are 
seventy 
sky positions 
that kill the spin-0 modes in the null streams. 
If a GW source with an electromagnetic counterpart is found 
in one of the 
seventy 
sky positions, 
the spin-1 modes will be testable separately from 
the spin-0 modes by the null stream method. 
We examined also a superposition of the two null streams 
to show that any one of the three modes 
(one combined spin-0 and two spin-1 modes) 
can be eliminated by suitably adjusting a weighted superposition of the null streams 
and thereby a set of the remaining polarization modes can be 
experimentally tested.

It is left for future work to perform a more comprehensive study 
for instance by taking account of expected noise at each detector of 
the near-future network HLVK.

\begin{acknowledgments}
We wish to thank Hideyuki Tagoshi, Yousuke Itoh and Nobuyuki Kanda 
for fruitful conversations at the first 
annual symposium (held at ICRR of University of Tokyo, 5th-7th March 2018) 
of the innovative area 
"GW Physics and Astronomy: Genesis (PI: Takahiro Tanaka)". 
We would like to thank 
Yuuiti Sendouda, Yuya Nakamura and Ryunosuke Kotaki 
for the useful conversations. 
We thank Atsushi Nishizawa, Seiji Kawamura, Shinji Miyoki and Masaki Ando 
for the useful comments on the first version of the manuscript. 
This work was supported 
in part by Japan Society for the Promotion of Science (JSPS) 
Grant-in-Aid for Scientific Research, 
No. 17K05431 (H.A.), 
and 
in part by Ministry of Education, Culture, Sports, Science, and Technology,  
No. 17H06359 (H.A.).  
\end{acknowledgments}

\newpage

\clearpage

\newpage

\newpage


\begin{thebibliography}{99}
\bibitem{footnote0}
In this paper, we introduce a notation as 
$V\equiv V_x$ and $W\equiv V_y$ 
for vector-$x$ and vector-$y$ modes, respectively, 
in order to avoid notational confusions for readers.
\bibitem{footnote2}
This is true for current ground-based detectors. 
However, it is not true for, e.g. planned Einstein Telescope or LISA, 
where the arms are not at 90 degrees with respect to each other.
\end{thebibliography}
\end{document}